\documentclass[preprint]{aastex}
\begin{document}
\title{The Unevolved Main Sequence of Nearby Field Stars and the Open Cluster Distance Scale}
\author{Bruce A. Twarog, Barbara J. Anthony-Twarog, and Frederika Edgington-Giordano}
\affil{Department of Physics and Astronomy, University of Kansas, Lawrence, KS 66045-7582}
\affil{Electronic mail: btwarog@ku.edu, bjat@ku.edu, fred82@ku.edu}

\begin{abstract}
The slope and zero-point of the unevolved main sequence as a function of metallicity are investigated using a
homogeneous catalog of nearby field stars with absolute magnitudes defined with revised $Hipparcos$ parallaxes,
{\it Tycho-2} photometry, and precise metallicities from high-dispersion spectroscopy. $(B-V)$ - temperature
relations are derived from 1746 stars between [Fe/H] $= -0.5$ and +0.6 and 372 stars within 0.05 dex of solar abundance; 
for T$_e$ = 5770 K, the solar color is $B-V$ = 0.652 $\pm$ 0.002 (s.e.m.). From over 500 cool dwarfs between [Fe/H] =
$-0.5$ and +0.5, $\Delta(B-V)/\Delta$[Fe/H] at fixed $M_V$ = 0.213 $\pm$ 0.005, with a very weak dependence upon the
adopted main sequence slope with $B-V$ at a given [Fe/H]. At Hyades metallicity this translates into $\Delta M_V/\Delta$[Fe/H] at fixed 
$B-V$ = 0.98 $\pm$ 0.02, midway between the range of values empirically derived from smaller and/or less homogenous
samples and model isochrones. From field stars of similar metallicity, the Hyades ([Fe/H] =
+0.13) with no reddening has $(m-M)_0$ = 3.33 $\pm$ 0.02 and M67, with E$(B-V)$ = 0.041, $A_V$ = 3.1E$(B-V)$, and [Fe/H] = 0.00, has
$(m-M)_0$ = 9.71 $\pm$ 0.02 (s.e.m), where the errors quoted refer to internal errors alone. At the extreme end
of the age and metallicity scale, with E$(B-V) = 0.125 \pm 0.025$ and [Fe/H] $= +0.39 \pm 0.06$, comparison of the fiducial relation
for NGC 6791 to 19 field stars with $(B-V)$ above 0.90 and [Fe/H] = +0.25 or higher, adjusted to the metallicity
of NGC 6791, leads to $(m-M)_0 = 13.07 \pm 0.09$, internal and systematic errors included. 
\end{abstract}

\keywords{Galaxy: open clusters and associations: individual (Hyades, M67, NGC 6791)}

\section{INTRODUCTION}

Distance determination to open and globular clusters is key to placing them in the proper Galactic evolutionary 
context and an indispensable component in evaluating stellar evolution as a function of mass, chemical composition, 
and age. For nearby clusters like the Hyades, Praesepe, the Pleiades and Coma, $Hipparcos$ parallaxes \citep{esa}, 
coupled with proper-motion and radial-velocity memberships, have generated precise distances, independent
of the cluster's composition and reddening, though these results have not been without controversy 
\citep{pi98, vl99, sn01, ma02, so05, vl09}. Any remote cluster 
beyond the reach of parallax with a known chemical makeup comparable to a nearby cluster can be compared differentially using stars on 
the unevolved main sequence to obtain a reliable distance (see, e.g., \citet{pi04, an07}). While unevolved main sequences of 
nearby clusters are ideal reference points due to the uniformity of age and composition among the cluster stars, the range
in chemical composition sampled by the nearby clusters is approximately [Fe/H]$ = -0.2$ to +0.2. For open clusters outside this range and
for all globular clusters, a traditional fallback procedure is to use cooler field dwarfs with parallaxes and well-defined
abundances either to isolate a stellar sample of similar abundance or to interpolate among stars that bracket the metallicity of interest 
\citep{tw99, gr03}. Given the broad gaussian distribution in [Fe/H] for field stars, this approach becomes more of a challenge for
clusters whose metallicity places them in the extended, low-metallicity tail of the field star sample. 

A routine alternative to field star comparison has been the construction of theoretical isochrones which can be tuned to 
any combination of composition and age, though ultimately these models must be linked to the empirical data of the field stars 
by matching the theoretical model combinations of mass, luminosity, composition and temperature to the observed values of mass, 
absolute magnitude, abundance and color at a given age. For stars of approximately solar mass and higher, there is reasonable agreement 
among the various isochrone compilations currently available in the literature and on-line, with most discrepancies among the 
models understandable in terms of differences in the adopted parameterizations of the internal physics, the model atmospheres, and 
the relations used to transfer from the theoretical to the observational plane \citep{pad, bas04, de04, vrs06, ma08, des08}. 
As one might expect, the discrepancies within the observational color-magnitude diagram (CMD) plane grow larger as one moves 
to stars of lower mass and/or more extreme compositions, reflecting the same paucity of empirical constraints found when 
attempting a direct match of distant clusters to nearby field stars with parallaxes.

Issues of reddening and metallicity determination aside, the more contentious discussions of cluster distances generally 
have focused on metal-poor systems of the thick disk and halo, while the metal-rich end of the distribution has been moderately 
immune due to the rich sample of nearby stars of solar abundance, the proximity of the Hyades, and the rarity of clusters with 
compositions significantly higher than the Hyades. For almost 50 years the one potential exception has been the 
old open cluster, NGC 6791, though the extent of its anomaly has been hidden beneath disagreements over its reddening, 
composition, distance, and age. In recent years significant progress has been made in constraining the first two parameters, both of 
which are critical to defining the third using main-sequence fitting, while the fourth has no bearing on the distance if 
the comparison is made to stars of sufficiently low mass. In every instance where revised reddening and/or metallicity estimates 
have been obtained, an improved distance modulus has been derived through comparison with theoretical isochrones 
\citep{bed08, ca06, ca05, ki05, stet, ch99, kr95, tr95, ga94, mj94, de92}. For the apparent distance modulus, the current 
range from main sequence fitting extends from a low of 13.1 \citep{stet} to a high of 13.6 \citep{slv03,atm07}. While the scatter 
is partly due to the adoption of different values for [Fe/H] and E$(B-V)$, a large portion is tied to disagreement over the location of 
the unevolved main sequence among the isochrones at high metallicity and the range of the CMD used to define an adequate fit to 
the models; in some instances, the unevolved main sequence and giant branch features are used simultaneously to optimize the fit. 

To circumvent the issues presented by the discrepancies among metal-rich isochrones, it was decided that the distance to NGC 6791 
could best be obtained by an empirical fit to unevolved field stars of similar composition. The feasibility of this option has been 
enhanced by the more restricted range among recent determinations of the cluster parameters, the availability of revised parallaxes 
and broad-band photometry from {\it Hipparcos/Tycho-2} \citep{esa,vl07}, and the compilation of a catalog of precise spectroscopic 
abundances, temperatures, and surface gravities on a common scale for almost 2100 nearby field stars. While the initial motivation 
for this study was the distance to NGC 6791, the need to define the precise location of the unevolved main sequence at high metallicity 
using the very restricted field star sample at comparable [Fe/H] generated a more comprehensive investigation of the metallicity 
dependence of the unevolved main sequence for stars of typical disk metallicity ([Fe/H] = -0.5 to +0.5). This paper builds upon 
the approach laid out in \citet{pe03}, using a sample expanded by an order of magnitude to derive the change in $M_V$ for cool dwarf 
stars at a given $B-V$ as [Fe/H] is varied from -0.5 to +0.5. While it is usually assumed that the ratio, $\Delta M_V/\Delta$[Fe/H], 
is constant with metallicity for unevolved disk dwarfs, derived values from observation and theoretical models vary 
by a more than a factor of two \citep{ko02, pe03, pi04, ka06}. The goal of this investigation is to explore the assumption of a
constant ratio, derive it, and test the consistency of the field star main sequence when applied to well-studied open clusters. Unless 
noted otherwise, errors quoted are standard deviations.
 
The layout of the paper is as follows. In Sec. 2 we will describe the database of astrometry, photometry, and spectroscopy for 
the nearby stars that is used in Sec. 3 to define the characteristics of the unevolved main sequence as a function of color and metallicity. In Sec. 4, 
the derived CMD relation is used to estimate distances to the well-studied open clusters, M67 and the Hyades, as well as the
extremely metal-rich cluster, NGC 6791. Sec. 5 contains a summary of our conclusions. 

\section{THE FIELD STAR SAMPLE}
\subsection{Fundamental Properties - The Metallicity Scale}
Delineation of the unevolved main sequence as a function of metallicity and color requires a significant sample of stars with reliable
abundance estimates, homogeneous and precise photometry that can be compared to the cluster data, and accurate absolute magnitudes. 
The sample of choice to fulfill the first condition is the spectroscopic catalog of abundances compiled by \citet{tva07},
expanded by the inclusion of three published spectroscopic surveys that have appeared since 2007. The core dataset that defines the
metallicity, temperature and surface gravity scale of the entire catalog is that of \citet{vf}, which supplies half of the stars in
the final sample. The addition of three new sources of spectroscopic data \citep{fuh08, sou08, mis08} raises the total source number to 29 and the
number of stars in the current catalog to 2085. While half of the stars come from the catalog of \citet{vf}, that 
database has been doubled without any significant reduction in the quality of the composite [Fe/H] measures. For details on the selection 
of the literature sources and the merger process, the reader is referred to \citet{tva07}. For the three recent additions, the numbers of stars
used in the transformations and the adopted errors for a single abundance estimate are 228 and 0.025 for \citet{sou08}, 67 and 0.031 for 
\citet{fuh08}, and 87 and 0.060 for \citet{mis08}, respectively.

While a concerted effort has been made to ensure the homogeneity of the abundances obtained by merging data from multiple sources,
for the purpose of probing potential differences with other derivations of the position of the unevolved CMD with [Fe/H], it is useful
to explore how the metallicity scale of \citet{vf} compares to that of other large abundance surveys. The two primary comparisons of interest
are to the photometrically-based abundance estimates of \citet{no04}, updated in \citet{hol07}, and the comprehensive spectroscopic 
catalog compiled by \citet{tay05}. For both sources, our spectroscopic data were cross-matched with the on-line catalogs; stars were retained 
if the published $T_e$ was between 4500 K and 6500 K, with absolute differences in [Fe/H] below 0.45 dex. For \citet{hol07} this produced 
1519 stars in common while the comparison dataset from \citet{tay05} was 898 stars. In each case, residuals in the sense (CAT-this work) were 
calculated and polynomials fit to the data as a function of the catalog [Fe/H] using a linear-least-squares routine. Due to the small number 
of metal-poor stars in the current catalog, the metallicity range was limited to stars with [Fe/H] above $-1.0$. Linear and quadratic fits were 
made to the entire sample and to the data sorted by $T_e$ into four bins 500 K wide between 4500 K and 6500 K. In each case, initial fits were 
made and, after eliminating points more than three standard deviations away from the mean, repeated. The results were extremely consistent for 
both catalogs; inclusion of quadratic terms failed to have a statistically significant impact upon the quality of any fit and were eliminated. 
A similar conclusion was confirmed regarding the removal of potentially deviant points, a byproduct of the large number of stars populating 
every temperature bin except the coolest, where the numbers ranged from 34 to 43. For the coolest bin, no stars were ever rejected as being too deviant.

As an initial comparison, the average residuals and their standard deviations are -0.044 $\pm$ 0.061 from 871 stars \citep{tay05} and -0.072 $\pm$ 0.084
from 1487 stars \citep{hol07}, implying that the \citet{vf} system is slightly more metal-rich than the others. The differences among the
catalogs become somewhat more distinct if one adds a linear term to the comparison. For \citet{tay05}, the residuals for 871 stars follow the relation

\medskip
\centerline{$\Delta {\rm[Fe/H]} = -0.036 (\pm 0.003) + 0.044 (\pm 0.008) {\rm[Fe/H]_{TAY}}$}
\medskip
\noindent
with a dispersion of 0.060 dex. For \citet{hol07}, the comparable relationship for 1482 stars is

\medskip
\centerline{$\Delta{\rm[Fe/H]} = -0.084 (\pm 0.002) - 0.121 (\pm 0.007){\rm [Fe/H]_{HOL}}$}
\medskip
\noindent
with a dispersion of 0.078 dex. The good agreement with the data of \citet{tay05} is encouraging but not surprising since the
majority of the spectroscopic surveys included in our compilation are included in the earlier comprehensive analysis. However, the catalog
of \citet{tay05} does not include the data of \citet{vf}, the defining catalog of our metallicity scale, so the agreement does imply that 
the two metallicity scales are very similar and transformations from one sample to the other are straightforward. The relation also 
demonstrates that the metallicity range among the disk stars in the catalog of \citet{tay05} is slightly larger than we would claim, but
the two systems are in closer agreement at the metal-rich end of the scale. By contrast, the abundance catalog of \citet{hol07} exhibits
a disk metallicity range reduced by one-sixth compared to that of \citet{tay05}, with the absolute discrepancy growing larger at higher metallicity.
Thus, a star with [Fe/H] = +0.5 in our catalog would have [Fe/H] = +0.36 on the system of \citet{hol07}. 

Finally, if one looks at the trends with temperature, there are only minor differences in the linear relations for the data of \citet{tay05}; the
dispersion in the residuals and the number of stars as one goes from the 6500 K bin to the 5000 K bin are (0.045, 270), (0.060, 416),
(0.070, 141), and (0.068, 43), respectively. By contrast, the comparable data for \citet{hol07} exhibit increasing scatter among the cooler
bins, (0.068, 462), (0.070, 756), (0.104, 234), and (0.127, 34), respectively, as well as a significant increase in the slope of the linear 
relation for the coolest stars. Photometric abundances for stars with T$_e$ below 5000 K on the \citet{hol07} scale are questionable.   

\subsection{Fundamental Properties - Parallax and Color}
Of the 2085 stars in the primary catalog, 2043 have parallaxes available in the revised catalog of $Hipparcos$ \citep{vl07}; the metallicity distribution
of this sample is shown in Fig. 1. While 86 stars are more metal-rich than [Fe/H] = +0.30, most are only slightly so; half of the stars fall 
between [Fe/H] = 0.31 and 0.35. Moreover, preliminary analysis demonstrated that the majority  of the stars fall in the vertical turnoff region 
for stars of the old disk, making them inappropriate for comparison with stars on the unevolved main sequence. 
For the purpose of deriving the distance to an exceptionally metal-rich cluster, the small sample could be expanded to include stars with 
[Fe/H] as low as +0.25, if reliable metallicity-dependent shifts in the CMD could be generated for unevolved main sequence stars. 

To derive metallicity-dependent CMD shifts, the field stars must be on a common photometric system that is readily transformable to
a significant cluster sample reaching the unevolved main sequence; for now, this restricts the analysis to broad-band $BV$ data. 
Fortunately, fewer than 1.5$\%$ of the stars with spectroscopic abundances and $Hipparcos$ parallaxes lack reliable $BV$ photometry on 
the {\it Tycho-2} system \citep{hog00}; the exceptions are dominated by stars that are too bright to have been observed 
photometrically or too faint to have reliable photometry. The {\it Tycho-2} $BV$ photometry \citep{hog00} has been converted to the 
Cousins-Johnson system using the transformation relations of \citet{mm02}, derived by applying polynomial fits to a table of mean 
points compiled by \citet{be00}. The table of \citet{be00} was constructed by a direct comparison of the original {\it Tycho} photometry with 
E-region standards composed primarily of B-G dwarfs and K-M giants. The revised relations exhibit significant structure as a function 
of $B-V$ in comparison with the original linear relations provided by the $Hipparcos$ catalog. The newer {\it Tycho-2} photometry follows 
the same relations defined by \citet{mm02}, but with higher internal precision.

A systematic error in the zero-point of the photometry and/or the slope of the unevolved main sequence could cause systematic shifts 
in the distance scale. Two basic tests of the reliability of the transformed {\it Tycho-2} data are available. First, \citet{pe03} defined the
approach of interest in this study to constrain the distance scale for nearby open clusters studied by $Hipparcos$. To ensure a reliable 
photometric comparison between the clusters and field stars, \citet{pe03} obtained $BVRI$ photometry of 54 cool field dwarfs over a 
range in [Fe/H] from $-0.56$ to +0.44, as defined by intermediate-band photometry recalibrated to avoid the issues raised by \citet{tw02}. 
The broad-band photometry is on the Cousins E-region system; the $V$ magnitudes and $B-V$ indices have mean standard deviations 
of $\pm$0.007 and $\pm$0.004 mags, respectively. We have taken the {\it Tycho-2} $BV$ photometry of these 54 stars and processed it 
through the same transformation relations as our catalog stars. For 54 stars, the mean residuals in $B-V$ and $V$, in the 
sense (converted Tycho - PER), are +0.0007 $\pm$ 0.0203 and +0.0029 $\pm$ 0.0301. In $B-V$, if the 2 stars with absolute 
residuals above 0.04 mags are excluded, the mean for the 52 stars becomes +0.0003 $\pm$ 0.0166. In $V$, if the two stars with residuals
in $V$ greater than 0.05 mags are removed, the mean for the 52 remaining stars becomes +0.0012 $\pm$ 0.0215. Note that in all
cases the errors are dominated by the uncertainty in the {\it Tycho-2} photometry.

Second, for reasons that will become apparent in Sec. 4, we will compare the $BV$ photometry for stars in the Hyades with the 
transformed {\it Tycho-2} data for the same stars. \citet{joh55} was selected as the primary source of broad-band cluster data. After
eliminating composite systems and variable stars, 100 stars with $B_T-V_T$ redder than 0.4 remained. For these 100 stars, the mean residuals
in $B-V$ and $V$ are $-0.011$ $\pm$ 0.028 and +0.017 $\pm$ 0.028, respectively. The offset in color, implying that the \citet{joh55} 
photometry is too blue, is consistent within the errors with the derived offsets of $-0.009$ $\pm$ 0.001 (s.e.m.) mag and $-0.006$ $\pm$ 0.004 (s.e.m.) 
in $B-V$ found by \citet{an07} in comparisons with the photometry of \citet{jon06} and {\it Tycho-2} data, respectively. The respective comparisons in 
$V$ indicate that the \citet{joh55} data are too faint by +0.026 $\pm$ 0.002 (s.e.m.) and +0.018 $\pm$ 0.003 (s.e.m.), again in good agreement with
the results derived above. The $B-V$ color offset between the Hyades and the {\it Tycho-2} system has also been confirmed by the analysis of \citet{jon08}.

To test for a color dependence among the residuals, we have applied a zero-point offset of +0.011 to the residuals for the Hyades and merged
the two samples. Fig. 2 shows the trend with color for 154 stars; filled squares are the data of \citet{pe03} while circles are the
Hyades observations. The error bars for each point are based upon the combined errors in both measures, but again are dominated
by the errors in the {\it Tycho-2} data. The apparent lack of a slope is confirmed by attempts to fit a line through the data; the resulting slope 
remains decidedly zero whether the data are included with or without weighting by the inverse square of the uncertainties.    

The final element in the field star database is the absolute magnitude. Of the 2043 stars with parallax, 61 have parallax errors 
larger than 12.5\% of the parallax; these have been excluded from the sample. For the remaining 1982 stars, distances and $M_V$ have been 
calculated from the parallax with the inclusion of Lutz-Kelker corrections \citep{lk73}. Because of the small uncertainties among the
parallaxes, the average Lutz-Kelker shift is below 0.03 mag. Of the parallax stars, 17 are in composite systems or variable, while 
25 do not have broad-band colors from {\it Tycho-2}, leaving a final sample of 1940 stars with complete data. 

\section{THE $M_V$ - $(B-V)$ - [Fe/H] RELATION}
\subsection{The $(B-V), T_e$ Relations}
While our primary interest is the CMD location of unevolved stars as a function of color and metallicity, it is useful to explore 
first the relationship between effective temperature and color. The temperature scale of the stars is tied to that
of \citet{vf}, derived from a comparison of high-dispersion spectra to a grid of synthetic spectra constructed using the models of \citet{ku92}
and interpolated to optimize the match between observation and theory. By defining the relation between temperature and color, we can use 
the spectroscopic temperatures to generate $B-V$ for stars with no $B-V$ photometry or for stars with
excessively large photometric errors, a not uncommon issue for the coolest stars in the sample. Additionally, the color-temperature 
relationship will allow a direct comparison between the current system and that adopted by others using different spectra, analyses, and
photometry. 

As a simple preliminary test, we have isolated the 375 stars in the final sample with [Fe/H] between $-0.05$ and +0.05 and $T_e$ between 4250 K and 7150 K. 
Excluding three stars with residuals between the calculated and observed $B-V$ greater than 0.10 mag, a fit between $T_n$, 
defined as ($T_e$ - 5770)/5770, and $B-V$ weighted by the inverse square of the uncertainty in $B-V$ for the remaining 372 stars produces:

\medskip
\centerline{ $B-V = 0.651 (\pm 0.001) - 1.966 (\pm 0.016) T_n  + 2.079 (\pm 0.140) T_n^2$}
\medskip
\noindent
Inclusion of a cubic term does not lead to a statistically significant improvement in the fit. The uncertainty estimates for $B-V$ included the 
photometric uncertainty and the dispersion in $T_e$, translated to an error in $B-V$ and combined in quadrature with the photometric error.
The standard deviation among the residuals in $B-V$ is 0.021 mag. Since the solar temperature of the models used to derive the spectroscopic
abundances is set at 5770 K \citep{vf}, the 
implied $B-V$ color of the Sun for this combination of stellar parameters and photometry is $B-V$ = 0.651.

Next, the stellar sample was sorted in bins 0.1 dex wide ranging from [Fe/H] = +0.55 to $-0.95$ and a quadratic polynomial fit was made for each
bin using the same temperature parameterization as above. For these fits stars were included if they fell in the temperature range from $T_e$ = 4000 K 
to 7500 K; points were weighted in the same manner as above.  Fig. 3 illustrates the morphology of the mean relations plotted 
differentially relative to the color relation for solar metallicity, in the sense ($B-V$)$_{[Fe/H]}$ - ($B-V$)$_{0.00}$. 
For clarity, mean relations are plotted for every other metallicity bin starting with [Fe/H] = +0.40 at the top. The curves are 
plotted only over the $T_e$ range covered by the sample used to define the curve. The relations exhibit reasonable morphological 
consistency between [Fe/H] = +0.5 and -0.5, with the predominant change being a regular shift in the zero-point of the curves as 
the metallicity declines. However, the differential relations for stars with [Fe/H] below -0.5 exhibited significant deviations from 
linearity, as illustrated by the bottom curve in Fig. 3. It should be emphasized that this is not a product of small number statistics; while 
the color range is declining among the lower metallicity stars, the bins centered on [Fe/H] = $-0.5$, $-0.6$, and $-0.7$ contained 59, 54, 
and 34 stars, respectively. Whether this is an effect tied to real changes in the color-temperature relations at lower [Fe/H] or an 
artifact of the $T_e$ merger process for stars of low metallicity remains unknown and, for our purposes, irrelevant. To define our 
final color-$T_n$ relation, we have restricted the metallicity range to [Fe/H] = +0.6 to $-0.5$, including only stars with $T_e$ between 
4200 K and 7000 K. Using $T_n$ as defined above, polynomial fits to $B-V$ were made to 1759 points using multiple 
combinations of $T_n$ and [Fe/H]. As expected from the earlier discussion, no terms above a quadratic in T$_n$ were found to be 
statistically significant. For [Fe/H], only the linear term and no cross terms involving [Fe/H] and T$_n$ were retained. 
After removing 13 stars with final residuals in $B-V$ greater than 0.10 mag, the remaining 1746 stars produce the following 
color - $T­_n$ relation:

\medskip
\centerline{$B-V = 0.653 (\pm 0.001) - 1.995 (\pm 0.008) T_n + 1.933 (\pm 0.076) T_n^2 + 0.148 (\pm 0.003) {\rm [Fe/H]}$}
\medskip
\noindent
The solar color is found to be $B-V$ = 0.653, slightly larger than from the solar metallicity sample alone, but the same within the errors; the 
standard deviation among the residuals in $B-V$ is again 0.021 mag. 

\subsection{The $M_V$, $(B-V)$ - [Fe/H] Relation}
Unlike a cluster CMD, the field star sample at a given metallicity is composed of stars with a wide range in age and therefore an increasing
spread in $M_V$ at higher temperatures. To minimize the impact of age, the obvious solution is to select stars of low enough mass and
temperature to ensure that evolution has a negligible impact. For stars with metallicity as high as NGC 6791, the predicted turnoff 
color for an adopted reddening of E$(B-V)$ = 0.15 is near $(B-V)_0$ = 0.75; a reddening at 0.10 implies $(B-V)_0$ near 0.80. Therefore, 
to define the unevolved main sequence at high [Fe/H], we need to restrict our sample to stars redder than $B-V$ = 0.80. As the metallicity declines, 
this boundary will shift to bluer colors. The down side of this restriction is the increasing fraction of stars with larger errors in observed $B-V$ 
among the cooler dwarfs. To illustrate the effect, we plot in Fig. 4 the CMD for all stars between [Fe/H] = +0.05 and $-0.05$, including errorbars 
in $M_V$ from parallax measurements and in $B-V$ from the published photometric uncertainties. While the mean relation with increasing $B-V$ 
is well-defined, the fraction of stars with photometric scatter above 0.05 mag also increases. 

By contrast, we duplicate the same data in Fig. 5 using $B-V$ based upon the transformation between the spectroscopically 
determined $T_e$ and 
$B-V$ as derived in the previous section. Note that this more general transformation includes a small adjustment for the individual values of [Fe/H]. 
The error bars in $B-V$ are now based upon the calculated uncertainties in $T_e$. For the CMD in the range of interest, $B-V$ redder 
than 0.85, the scatter in color is reduced with no apparent change in the mean relation among the stars along the lower main sequence. Excluding
stars with color residuals above 0.10 mag, for 144 stars redder than $B-V$ = 0.72, the weighted average of the color residuals, in the sense
($(B-V)_{OBS} - (B-V)_{PRE}$), is +0.006 $\pm$ 0.009. Drawing the cut at $B-V$ = 0.90 reduces the sample to 62 stars with an average residual
of $-0.002$ $\pm$ 0.021. Because the scatter in $B-V$ based upon the converted $T_e$ is inherently smaller 
than the scatter among the observed colors and because it allows us to include in the CMD analysis stars that do not have reliable, if any, 
photometric measures on the {\it Tycho-2} $BV$ system, we will use the transformed $T_e$ estimates as our $B-V$ source for the CMD.

As a first probe of the unevolved main sequence relation, we have attempted a polynomial fit using only the solar metallicity data of Fig. 5. The initial
color range was restricted to $B-V$ = 0.72 to 1.30 and stars with excessive deviations from the mean trend were eliminated. The preliminary cut
adopted was 0.50 mag, but this was reduced to 0.35 mag after an initial iteration of the fit. Of 132 stars that initially met the color criterion, 
4 were eliminated from the final fit. Tests of the polynomial fit showed that the terms beyond the linear were statistically irrelevant. The final
main sequence relation for the solar metallicity stars is

\medskip
\centerline{$M_V = 1.75 (\pm 0.04) + 4.77 (\pm 0.05) (B-V)$}
\medskip
\noindent
The dispersion in the residuals about the mean relation is $\pm 0.13$ mag. If we extrapolate the relation to an assumed solar color of $B-V$ =
0.652, $M_V$ becomes 4.86, only slightly fainter than our adopted value of $M_{V_{\odot}}$ = 4.84. Given the age of the Sun, one might expect a larger
differential compared to the unevolved main sequence. We will return to this issue in Sec. 4. The linear fit is shown as a solid line in Fig. 5.

To derive the unevolved main sequence relation for a range of [Fe/H], our initial sample of 1982 stars with reliable abundances and parallax measures 
was restricted to single, non-variable stars with [Fe/H] between $-0.50$ and +0.50, $M_V$ fainter than +4.5, $B-V$ bluer than 1.30, and $B-V$ 
redder than $B-V = 0.2{\rm [Fe/H]} + 0.72$, cutting the dataset to 533 stars. Binaries were tagged from the high-dispersion spectra
obtained to derive the abundances in each study included in the composite catalog, while significant variability was checked through the Tycho 
catalog. After a preliminary polynomial fit to the data, including tests of up to 
cubic terms in $B-V$, quadratic terms in [Fe/H], and multiple cross-term combinations, only the linear terms in $B-V$ and [Fe/H] survived, the 
former result being consistent with what was found for the solar sample. All stars with residuals in $M_V$ greater than 0.35 mag were excluded 
and the final calibration repeated. The final sample consisted of 501 stars with a standard deviation among the residuals in $M_V$ of 0.135 mag. 
The derived polynomial function is:

\medskip 
\centerline{$M_V = 1.55 (\pm 0.05) + 5.00 (\pm 0.03) (B-V) - 1.07 (\pm 0.02) {\rm [Fe/H]}$}
\medskip
\noindent
For solar metallicity stars, the slope of the unevolved main sequence based upon the derived relation is somewhat steeper than found
previously, leading to stars that are 0.03 mag brighter at $B-V$ = 0.75, but 0.09 mag fainter at $B-V$ = 1.25. If we demand that the 
main sequence relation have the same slope at solar metallicity as derived from the solar sample, the relation becomes

\medskip
\centerline{$M_V = 1.75 (\pm 0.01) + 4.77 (B-V) - 1.04 (\pm 0.02) {\rm [Fe/H]}$}
\medskip
\noindent
The standard deviation among the 502 residuals increases slightly to 0.137. For reasons that will be discussed in Sec. 4, it is probable
that even the slope of 4.77 is too steep for the true, unevolved main sequence.

\subsection{Comparison to Previous Determinations}
As stated in Sec. 1, a primary goal of this investigation is to test the sensitivity of the absolute magnitude
of unevolved cooler main sequence stars as $B-V$ and [Fe/H] are varied. Qualitatively, our data relations confirm that, within the
current uncertainties, there is no statistically significant evidence for a variation in the ratio, $\Delta M_V/\Delta$[Fe/H],
with $B-V$ or [Fe/H] between +0.5 and $-0.5$ for truly unevolved stars. The constant ratio is consistent with the work of \citet{pe03},
using intermediate-band photometric abundances for an order-of-magnitude smaller sample, and the prediction of stellar isochrones used to
define the cluster distance scale as detailed in \citet{pi04} and \citet{an07}, but disagrees weakly with the analysis of a comparable databset of
hotter stars using abundances derived from $UBV$ indices \citep{ka06}. However, the scatter in the results becomes apparent
when the specific values of the slope are compared.

Because the technique adopted is an expanded version of the technique laid out by \citet{pe03}, we will translate our change in $M_V$ with
[Fe/H] at a given $B-V$ to a change in $B-V$ with [Fe/H] at a given $M_V$ to allow a direct comparison with their result. Since the
main sequence slope at a given [Fe/H] is assumed to be linear, these comparisons are equivalent. From 54 stars, \citet{pe03} 
find $\Delta(B-V)/\Delta$[Fe/H] = 0.154, with a very weak sensitivity of the final value to the adopted main sequence slope. 
With a main sequence slope of 4.77, this translates into $\Delta M_V/\Delta$[Fe/H] = 0.73. For our two relations above, 
$\Delta(B-V)/\Delta$[Fe/H] $= 0.214$ and 0.218, respectively. Changing the fixed main sequence slope to 4.5 and 5.5 produces 
$\Delta(B-V)/\Delta$[Fe/H]$ = 0.217$ and 0.208, respectively, confirming the insensitivity of the color gradient to the adopted main sequence 
slope but clearly indicating that the value of \citet{pe03} underestimates the metallicity effect by 29\%. Note that because 
$\Delta(B-V)/\Delta$[Fe/H] is constant, $\Delta M_V/\Delta$[Fe/H] will vary directly with the adopted slope for the main sequence. 

At the other end of the scale, \citet{pi03, pi04} have constructed empirically-adjusted isochrones to use in defining cluster distances
through main sequence fitting. The isochrones are built upon the models of \citet{si00} and transformed to the observational plane using
a variety of $T_e$-color relations that are empirically adjusted to ensure an ideal match to the Hyades. \citet{pi04} include a comparison of
the impact of the various $T_e$-color relations on the absolute magnitude of a star at a given color as [Fe/H] is varied between $-0.3$ and
+0.2 over the color range $B-V$ = 0.40 to 1.0. For their models, the zero-points of the absolute magnitude-[Fe/H] relations vary 
with the $T_e$-color relation adopted, but the slopes are extremely consistent: $\Delta M_V/\Delta$[Fe/H]$ = 1.4$, 35$\%$ larger than derived in this investigation 
and almost double that found by \citet{pe03}. The equivalent $\Delta(B-V)/\Delta$[Fe/H] $= 0.29$. It should be emphasized that these
numbers are tied to a specific set of models. A check of the many available on-line sources for theoretical isochrones shows that among
the cooler dwarfs, there is significant variation in $\Delta(B-V)/\Delta$[Fe/H] from one set of isochrones to another and,
among some sets, the slope can be found to vary with both $B-V$ and [Fe/H].

The final comparison is with the recent work of \citet{ka06}. This study derives a new calibration of metallicity as a function of
the color excess, $\delta(U-B)_{0.6}$, and, using stars with $Hipparcos$ parallaxes on the original system, defines an 
$M_V$, $(B-V)$, $\delta(U-B)_{0.6}$ relation. The range in $B-V$ for the sample is bluer than ours, covering approximately $B-V$ = 0.4 
to 0.9, with evolved stars eliminated using a magnitude cut roughly parallel to the unevolved main sequence in the CMD. The absolute magnitude 
calibration is constructed and tested with a number of sample variations; relations are derived for $M_V$ as a function of $B-V$ and 
$\delta(U-B)_{0.6}$ using all stars with parallax and for $\Delta M_V$ as a function of $B-V$ and $\delta(U-B)_{0.6}$ using only Hyades 
dwarfs and field halo dwarfs. The uniform pattern among the various relations is that the slope, $\Delta M_V/\Delta$[Fe/H],
varies with both $B-V$ and [Fe/H]. Using the final differential calibration applied at $B-V$ = 0.70 between [Fe/H] = +0.39 and $-0.50$, the
average $\Delta M_V/\Delta$[Fe/H] = 0.63, smaller than the value found by \citet{pe03}. The result is even more extreme if the 
change is translated to a $B-V$ ratio because the average main sequence slope defined by the main sequence data of \citet{ka06},
weighted toward the bluer end of the main sequence, is 5.5, leading to $\Delta(B-V)/\Delta$[Fe/H]$ = 0.115$.

\section{The Open Cluster Distance Scale}
\subsection{The Hyades and M67}
Independent of the sensitivity of $M_V$ with [Fe/H] at a given $B-V$, does the absolute scale of the parallax sample generate plausible
distance moduli for well-studied clusters? The obvious initial test case is the Hyades. Because of the respectable number of stars with
a metallicity approximating that of the Hyades, we can follow an approach similar to that for the solar metallicity sample in the previous section.
All stars with [Fe/H] within $\pm$0.05 dex of the Hyades metallicity have been identified, adjusted in $M_V$ to the metallicity of the Hyades 
and then compared to the Hyades main sequence relation.

What is the Hyades metallicity? We have identified Hyades stars within our catalog through a comparison with two sources of Hyades
members. From 18 Hyades stars that overlap with the spectroscopic catalog of \citet{tay05}, [Fe/H] = 0.129 $\pm$ 0.029 on our system.
The average offset of $-0.031$ dex is essentially identical to the predicted value of $-0.032$ dex from the relation for the entire
catalog as derived in Sec. 2. From 10 stars that overlap with the catalog utilized by \citet{pi04}, [Fe/H] = 0.137 $\pm$ 0.040.
We will adopt [Fe/H] = +0.13 for the Hyades as defined by our spectroscopic scale. 

For the Hyades main sequence we have adopted the mean relation derived by \citet{pi04}, equivalent to a parallax-defined 
true modulus of $(m-M)_0$ = 3.33, with two modifications. As detailed in Sec. 2, the \citet{pi04} photometric scale exhibits small 
offsets relative to the converted {\it Tycho-2} $BV$ system that defines our sample. We have adjusted the \citet{pi04} Hyades mean relation 
to our scale by adding 0.009 mag in $B-V$ and decreasing $V$ by 0.020 mag. This is equivalent to shifting the original relation in 
$M_V$ at a given color by $-0.06$ mag.

Fig. 6 shows a plot of the residuals, in the sense $(M_{VHyades} - M_{VField})$, as a function of $B-V$ for parallax stars with [Fe/H]
between +0.08 and +0.18, adjusted in $M_V$ to [Fe/H] = +0.13 using $\Delta M_V/\Delta$[Fe/H]$ = 1.04$. A color-dependent asymmetry
among the residuals is obvious. On the negative side of the distribution, there is an extremely sharp edge to the residuals 
near $-0.15$ mag which should be indicative of the scatter in the absolute magnitudes due to the combination of parallax errors, photometric 
errors, and temperature errors translated into errors in $B-V$. On the positive side of the residuals, there is a clear increase in the 
range of the scatter as one moves from the red to the blue side of the plot. The asymmetry is real and has two primary sources. 
First, stars that are increasingly older than the Hyades will be increasingly brighter than the Hyades main sequence at a given color, with 
the degree of the offset decreasing for redder stars. Second, the color distribution of the parallax sample is more heavily weighted 
toward bluer stars. While we can set a boundary to the sample at the blue end to exclude stars that are more likely to show the 
effects of evolution, random errors in the temperature estimates, translated into random errors in $B-V$, will preferentially shift more 
of the blue stars toward the red than vice versa. The impact of this asymmetry will be to steepen the derived main sequence slope 
obtained by fitting a straight line to field star parallax data, as done in Sec. 3, unless the blue color limit for the sample is set 
well redward of the expected evolutionary cutoff. This effect, in part, provides a plausible explanation for why the main sequence
relation for solar metallicity stars, as derived in Sec. 3, passes so close to the position of the evolved Sun at solar $B-V$.
For the adjusted Hyades mean relation, a linear fit to the data between $B-V$ = 0.73 and 1.30 produces

\medskip
\centerline{$M_V = 1.87 (\pm 0.005) + 4.53 (\pm 0.025)(B-V)$}
\noindent
As expected from the trend in Fig. 6, the slope is shallower than found for the sample as a whole or for the solar sample. If we adopt the 
slope of the Hyades relation as the correct match to the unevolved main sequence, the revised relation for field stars as a function of
metallicity becomes

\medskip
\centerline{$M_V = 2.00 (\pm 0.01) + 4.53 (B-V) - 0.98 (\pm 0.02){\rm [Fe/H]}$}
\noindent
It should be emphasized that while the inclusion of stars that are too blue can have a significant impact on the derived slope of
the main sequence relation, the mean residuals in $\Delta M_V$ are only mildly affected. Using a cutoff of $B-V$ = 0.70, bluer than
the value of $B-V$ = 0.75 adopted for stars of Hyades metallicity in defining the main sequence relation in Sec. 3, and excluding 
stars with absolute residuals greater than 0.5 mag, 
the remaining 105 stars generate a weighted average $\Delta M_V$ of +0.017 $\pm$ 0.005 (s.e.m.). 
If we draw the color cut for the sample at $B-V$ = 0.90, the comparable offset is $-0.006$ $\pm$ 0.010 (s.e.m.) from
51 stars, confirming that the revised Hyades relation is an excellent match to the field stars of comparable metallicity. Note that
the consistency between the $M_V$ scales from the Hyades analysis by \citet{pi04} and the current field star sample is equivalent to 
stating that had we matched the unevolved Hyades apparent main sequence to our field stars of comparable metallicity, the
derived modulus would be 3.33, independent of the previous determination.

Before moving on to M67, the trend seen in Fig. 6 may supply an explanation for the significantly smaller ratio of $\Delta M_V$ with
[Fe/H] found by \citet{ka06}, 0.63 as opposed to 0.98. The turnoff-dwarf sample is identified by a star's position in the CMD relative to
a line that runs roughly parallel to the main sequence between $B-V$ = 0.3 to 0.65. For redder stars, the sample limit is defined by
a 12-Gyr isochrone with [Fe/H] = $-0.3$.  For the bluer stars, a fixed boundary in $M_V$ at a given color will allow a larger range of 
evolved turnoff stars to be included in the sample as [Fe/H] decreases. Thus, the shift in the average $M_V$ of the sample above the 
unevolved main sequence at a given color will increase as [Fe/H] decreases. Qualitatively, this should steepen the slope of the mean
main sequence with decreasing [Fe/H] while compressing the range in $M_V$ at a given color for a given range in [Fe/H], leading to
a smaller derived ratio of $\Delta M_V$ with [Fe/H]. 

As a second test of the revised main sequence relation, we have adjusted all parallax stars between [Fe/H] = $-0.05$ and 0.05 to an assumed
[Fe/H] = 0.0 and calculated residuals relative to the adjusted Hyades relation shifted to [Fe/H] = 0.0. Excluding all stars with absolute 
residuals greater than 0.5 mag, for $B-V$ greater than 0.72, 130 stars produce $\Delta M_V$ = +0.038 $\pm$ 0.005 (s.e.m.). Placing the 
color cut at $B-V$ = 0.90 reduces the offset to +0.033 $\pm$ 0.008 (s.e.m.) for 57 stars. The conclusion is that application of the mean 
relation at solar metallicity to a cluster with the same properties as the nearby solar field stars will lead to an apparent modulus 
that potentially is too small by 0.03 mag. With this caveat in mind, we can now derive the apparent modulus for the well-studied open cluster, M67.

Unlike the field and Hyades stars included in the current study, M67 is reddened and not tied directly to the spectroscopic
scale or the {\it Tycho-2} $BV$ system. Within the extensive literature dealing with the fundamental properties of M67, the two most relevant recent
studies are \citet{an07} and \citet{pa08}. The former builds upon the approach of \citet{pi03,pi04} to construct 
semiempirical isochrones in multiple colors to simultaneously define the reddening, metallicity, and distance for a number of open clusters,
including M67. The final values derived for M67 are E$(B-V)$ = 0.042, [Fe/H] = $-0.02$ and $(m-M)$ = 9.74. The latter paper uses high-resolution spectra 
and line-depth-ratios, H$\alpha$ lines, and Li measures to identify stars within M67 that resemble solar analogs. With a derived solar color 
of $B-V$ = 0.649, E$(B-V)$ = 0.041 and [Fe/H] = +0.01, \citet{pa08} find $(m-M)$ = 9.78. If we fix E$(B-V)$ at 0.041 and set [Fe/H] = 0.00,
the two investigations imply $(m-M)$ = 9.76 and 9.77, respectively. 

The last piece of the puzzle requires a link between the adopted $BV$ photometry for M67 \citep{sa04} and that of the
transformed {\it Tycho-2} colors. The only direct comparison is that of \citet{jon08} where the Hyades, M67, and transformed {\it Tycho-2}
data are coupled through SAAO observations, all ultimately tied to SAAO E-region standards. Within the uncertainties, the multiple
comparisons imply that the color offsets required to transform the \citet{joh55} Hyades $BV$ data to the {\it Tycho-2}/SAAO system are similar
to those applicable to the \citet{sa04} $BV$ data of M67 tied to the Johnson system through the standards of \citet{ste00} and \citet{lan92}. 
We therefore have added 0.009 mag to the $B-V$ indices of M67. For $V$, a direct link between the M67 photometry of \citet{sa04} and the 
{\it Tycho-2}/SAAO system is unavailable. Standards for the M67 CCD calibration are from to the systems of \citet{ste00} and \citet{lan92}; 
comparisons with other M67 sources indicate that the \citet{sa04} $V$ magnitudes are a match to this standard system at the $\pm$0.004 mag level. 
By default, no adjustment is made to the M67 $V$ photometry.

From the compilation of \citet{sa04}, we have selected all single-star probable members of M67 with adjusted and
reddening-corrected $(B-V)_0$ above 0.75. From 38 stars, giving all individual moduli equal weight, the average apparent modulus 
for M67 is $(m-M)$ = 9.807 $\pm$ 0.014 (s.e.m). With weighting based upon the inverse square of the uncertainty in $V$, including the 
effects of the color errors, the apparent modulus rises slightly to 9.820 with the uncertainty cut in half. If we were to fit the 
M67 main sequence directly to field stars of comparable metallicity, i.e., stars between [Fe/H] = $-0.05$ and 0.05 adjusted in $M_V$ to an 
adopted value of [Fe/H] = 0.00, the moduli would be increased by 0.033 mag, or $(m-M)$ = 9.84 for the unweighted average. With $A_V$ = 3.1E$(B-V)$,
this becomes $(m-M)_0$ = 9.71 $\pm$ 0.02 (s.e.m.), taking the uncertainty in the reddening into account. 

Is the offset between the current value for the modulus and those of \citet{an07} or \citet{pa08} significant? The errors quoted above are
the internal errors defined by the scatter within the photometry. Because the comparisons are made under the same assumptions for the
reddening and metallicity, at a systematic level, the dominant source of uncertainty remains the size and applicability of the color 
adjustment to transfer the $B-V$ system of \citet{sa04} to the {\it Tycho-2} system that defines the field stars and the Hyades. If 
we adopt $\pm$0.009 as a plausible estimate for the uncertainty in the differential color correction, this alone leads to an 
uncertainty in $(m-M)$ of $\mp$0.040. 

There is one potential source of a systematic offset between the distance scale of \citet{an07} and ours. The main sequences used 
to fit the cluster sequences are tied to theoretical isochrones empirically modified to match the Hyades at an adopted [Fe/H] = +0.13 
and $(m-M)_0$ = 3.33; the Hyades mean relation transferred to the {\it Tycho-2} system is an excellent match to
field stars of comparable metallicity in our database. When shifted in $M_V$ at a given $B-V$ using $\Delta M_V/\Delta$[Fe/H]$ = 0.98$,
the Hyades relation overshoots the field star solar sample by a modest amount, i.e., it is too faint compared to the field stars, implying
that our slope over this short distance is too large. However, as discussed previously, the isochrones employed by \citet{an07} predict 
$\Delta M_V/\Delta$[Fe/H]$ = 1.4$, though the trends with metallicity for individual clusters in \citet{an07} indicate values between
1.2 and 1.4. If the isochrone sequences defining the main sequence relations predict too great a change in $M_V$ with [Fe/H] relative to
the Hyades, for a shift in [Fe/H] of $-0.13$ dex and $\Delta M_V/\Delta$[Fe/H]$ = 1.25$, the main sequence matched to M67 will be 
systematically too faint by 0.035 mag relative to our system and 0.065 mag too faint compared to the field stars.

\subsection{The Distance to NGC 6791}
While the old open cluster, NGC 6791, isn't the primary focus of this investigation, it does have a valuable role to play as a test
of the more extreme limits of the field star main sequence calibration. M67 and the Hyades are both well-placed among the richly
populated metallicity distribution and differentially separated by only a modest shift in [Fe/H]. By contrast, current metallicity
estimates from photometric and spectroscopic techniques \citep{or06, ca06, gr06, ca07, atm07, bo09} generate a weighted average of
[Fe/H] = +0.39 $\pm$ 0.06. Moreover, while the exact age of the cluster remains uncertain due to questions regarding the
reddening and metallicity, the dereddened color of the stars within the vertical turnoff will lie between $(B-V)_0$ = 0.74 and 0.84
for E$(B-V)$ = 0.15 to 0.10. Minimally, one should choose only stars with $(B-V)_0$ redder than 0.90 to derive the distance to 
NGC 6791. For [Fe/H] $\ge$ 0.30, only 10 stars in our sample meet the criterion. Therefore, any attempt to test the cluster distance
scale at the metal-rich end will require a shift of field stars from lower [Fe/H] to generate a statistically significant sample.

Fig. 7 shows the residuals in a comparison between the field star data, adjusted to [Fe/H] = 0.39, and the shifted Hyades relation 
using $\Delta M_V/\Delta$[Fe/H]$ = 0.98$. Stars with absolute residuals above 0.50 mag have been excluded. Filled circles are 
stars with [Fe/H] above 0.29, while open circles are stars between [Fe/H] = 0.25 and 0.29. While the scatter is larger, the pattern 
that emerges is similar to that of Fig. 6. The bluer ($B-V$ below 0.90) stars are systematically brighter than predicted from the 
adjusted Hyades relation, as expected if these stars are exhibiting the effects of evolution away from the main sequence. By contrast, 
the redder stars approximately scatter around mean relation, though there may be weak evidence for a slope among the residuals, implying 
that the Hyades relation is too shallow compared to the very metal-rich field stars. We will return to this point in Sec. 5.

As an initial check, the 10 stars between [Fe/H] = 0.30 and 0.47 with $(B-V)_0$ above 0.9 have been compared to the adjusted Hyades 
relation and individual residuals calculated for each field star relative to the Hyades using weighting by the inverse errors in $M_V$, 
including the propagated errors from the uncertainty in the temperatures used to define the colors. For 10 stars, the average residual, 
in the sense ($M_{VHyades}$ - $M_{VField}$), is 0.002 $\pm$ 0.026 (s.e.m.). If we lower the color limit to $B-V$ = 0.85, the sample 
expands to 16 stars, but the mean residual increases significantly to 0.091 $\pm$ 0.018 (s.e.m.), a confirmation of the asymmetry in 
the distribution with decreasing color. If we keep the color limit at $B-V$ = 0.90, but lower the field star limit to [Fe/H] = 0.25, the 
mean residual is 0.004 $\pm$ 0.019 (s.e.m.) from 19 stars, indicating that the field stars over the color range
of $(B-V)_0$ = 0.9 to 1.1 shifted to the metallicity of NGC 6791 should predict the same distance modulus at the $\pm 0.02$ mag level as a 
direct comparison of the cluster to the adjusted Hyades relation over the same color range.

With the metallicity set, two key issues remain in establishing the comparison. Estimates of the cluster reddening have ranged from E$(B-V)$ = 0.08 to
0.26, but recent work based upon a variety of techniques has limited the plausible range to E$(B-V)$ = 0.10 to 0.15 \citep{atm07}. As
a compromise, we will adopt E$(B-V)$ = 0.125 $\pm$ 0.025. 

The second issue is the perennial question of the photometric system of the cluster $BV$ data. The most comprehensive and
internally reliable photometric sample available to date is that of \citet{stet} which is tied to the same standard star
network adopted by \citet{sa04} in the photometric calibration of M67. For internal consistency, we will apply the same
color offset of +0.009 mag to the NGC 6791 photometry as we did for M67 and the Hyades. No adjustment will be made to
$V$. For the cluster CMD relation, we will use the fiducial relation compiled by \citet{slv03} limited to an
adjusted $B-V$ below 1.25 since our reddest field star is at $(B-V)_0$ = 1.09 and the reddening is E$(B-V)$ = 0.125.

The results are illustrated in Fig. 8, which has the same symbol definitions as Fig. 7, using the fiducial relation
of NGC 6791 with $(m-M)$ = 13.45. For the same 19 stars redder than $B-V$ = 0.90, the average residual in $M_V$ is 0.005.
What is different compared to the Hyades comparison is the removal of the apparent asymmetry in residuals for bluer stars
and the weak evidence for a slope at redder colors. This change indicates that the fiducial relation for NGC 6791 is a
better match to the CMD position of the average field star of high metallicity than the Hyades, as one would expect if the
typical star in the sample is older than the Hyades cluster but younger than NGC 6791. The possibility that the
field stars may be younger than NGC 6791 comes from the sharp turn toward negative residuals for $B-V$ bluer than 0.85.
With the adopted cluster reddening of E$(B-V)$ = 0.125, this color marks the start of the rapid vertical
rise in the cluster turnoff, thereby leading to anomalously bright absolute magnitudes relative to less evolved stars.

For a fixed value of the reddening and metallicity, taking into account the potential uncertainty in the color offset applied
to the fiducial relation for NGC 6791, the apparent modulus is $(m-M)$ = 13.46 $\pm$ 0.049. \citet{slv03} supply no estimate
of the uncertainty in their fiducial relations. We have done an independent check on their relation by rederiving the fiducial
curve over the color range of interest using only stars in the core of NGC 6791. Stars between $B-V$ = 0.96 and 1.26 were
identified and retained if they fell within $\pm$0.30 mag of the \citet{slv03} relation. Stars between $B-V$ = 0.96 and
1.02  were adjusted in $V$ to the central color of the bin using the slope of the main sequence relation
as defined by \citet{slv03} and counted into bins 0.06 mag wide in $V$. The resulting histogram was fit with a gaussian
profile to define the peak of the distribution. The bin was then shifted redward in $B-V$ by 0.04 mag and the process
repeated. The fiducial points were compared to those of \citet{slv03}; the average difference for the 7 points, in the
sense (SLV - ours), is 0.009 $\pm$ 0.020, indicating that the \citet{slv03} fiducial relation is not a significant
source of uncertainty. If we allow for $\pm$0.025 uncertainty in E$(B-V)$ and $\pm$0.06 uncertainty in [Fe/H], the
combined internal and external errors imply $(m-M)$ = 13.46 $\pm$ 0.15, with the dominant source of error being the
reddening uncertainty. If the true modulus is calculated, the error bars are reduced because the reddening effect
is partially compensated by the correlated change in $A_V$; the true modulus is $(m-M)_0$ = 13.07 $\pm$ 0.09, internal and
external errors included. 

How does this result compare with current estimates? The definitive value for NGC 6791 at present is that of \citet{gru08}
based upon analysis of the cluster eclipsing binary, V20. The beauty of the technique is that the masses and radii can
be determined independent of the reddening and metallicity and age estimation can be carried out through comparison to
isochrones in the mass-radius plane without requiring a transformation of the theoretical parameters to the observational
plane. Conversion of the stellar parameters to luminosities and distances still requires an assumed reddening
and metallicity so that the observed stellar colors can be translated into effective temperatures and luminosities 
converted to $M_V$. Adopting E$(B-V)$ = 0.15 and [Fe/H] = +0.40, \citet{gru08} derive $(m-M)_0$ = 13.00 $\pm$ 0.10; if we 
had adopted the same reddening and metallicity, our result would have been $(m-M)_0$ = 13.13 $\pm$ 0.09.   

\section{Summary and Conclusions}
Distance determination for field stars and clusters remains a primary observational objective for those interested in
understanding stellar and Galactic evolution. Ideally, parallaxes for a large sample of nearby stars with well-defined
abundances and temperatures/colors would allow one to map the impact on the absolute magnitude of varying the abundance of a star
of a given temperature/color. With this information in hand, determining the distance to any cluster with reliable abundance and reddening
information becomes a straightforward task. Unfortunately, the reality is somewhat different. While reliable parallaxes are available
for a large sample of field stars and most have colors on the {\it Tycho-2} system, the critical component missing from the 
picture has been a comparable catalog of precise metallicity estimates. The observational approaches to defining the
ratio of $\Delta M_V$ with [Fe/H] \citep{ko02, pe03, ka06} have relied upon photometric abundances coupled to
a spectroscopic subset of the sample. The slope of the relation is either assumed to be constant with
[Fe/H] and color and/or tested through the use of theoretical isochrones. For \citet{ko02} and \citet{ka06}, the baseline used to
define the slope is extended to [Fe/H] below $-1$ through the inclusion of halo dwarfs and/or globular clusters. The
dataset of \citet{ka06} extends to stars where evolution off the main sequence is significant and may produce a
selection bias in the mean absolute magnitude with [Fe/H].

Starting with a dataset of almost 2000 stars with reliable parallaxes, spectroscopic abundances, and homogeneous
colors, we have reduced the sample to approximately 500 stars between [Fe/H] = $-0.5$ and +0.5 with colors red enough
that evolution off the main sequence should be negligible. However, as evidenced by the old, metal-rich
turnoff stars in NGC 6791, even our metallicity-dependent cutoff allows some significantly evolved stars into
the mix. Over the primary color range of interest, $B-V$ = 0.75 to 1.15, the ratio with metallicity, $\Delta M_V/\Delta$[Fe/H]$ 
= 0.98 \pm$ 0.02 with no evidence for a color or [Fe/H] dependence. Because this value assumes a universal slope
of 4.53 for the main sequence with $B-V$, it is probably better to define the relation in terms of 
$\Delta(B-V)/\Delta$[Fe/H]$ = 0.213 \pm$ 0.005, which is only weakly dependent upon the adopted slope of the main 
sequence. The concern that the slope of the main sequence with $B-V$ varies with [Fe/H] is real and not dependent
upon the multiple sets of theoretical isochrones which are inconsistent on this point. From the observed cluster sequences
analyzed in this investigation, the Hyades has a main sequence slope of 4.5; over the same color range, the fiducial
relation for NGC 6791 has a slope of 5.4.

With the field-star relations in hand, reliable estimation of cluster distances should follow. The challenge in this
phase is ensuring that the clusters are on the same photometric color and spectroscopic abundance system as the
field stars which, for more distant objects, also includes accurate reddening estimation. Because our catalog includes
Hyades stars and the Hyades is assumed to be reddening-free, the only potential source of controversy is the
photometric scale. Our analysis confirms what has been found by others \citep{jon06, an07, jon08}. The \citet{joh55}
photometric system in the Hyades is offset from the SAAO/E-region standards that define the transformed {\it Tycho-2}
system by approximately -0.009 mag and 0.02 mag in $B-V$ and $V$, respectively. When these corrections are applied to
the fiducial relations for the Hyades \citep{pi04}, the cluster produces an excellent match to the field stars at
the appropriate [Fe/H] with the cluster modulus set at $(m-M)_0$ = 3.33.

The second test of the system is a match to M67. Because this cluster and NGC 6791 are two steps further removed from the field star sample
in that there are no cluster stars with spectroscopic abundances in our catalog and the photometric data are not directly comparable
to the {\it Tycho-2} system, we have applied the same color offset found for the Hyades to the cluster data and made differential
comparisons based upon a commonly adopted reddening and metallicity estimate. If the M67 data of \citet{sa04} are matched to
the Hyades fiducial relation shifted to [Fe/H] = 0.00, the apparent modulus of the cluster becomes 9.81 $\pm$ 0.02; if a match
is made directly to the field stars of identical [Fe/H], the modulus increases by 0.03. The comparable value from \citet{an07} and \citet{pa08}
is 9.765. Note that part of the offset relative to \citet{an07} may be a product of the elegant but hybrid approach that defines
the main sequence relations using a mix of theoretical isochrones with color corrections defined by cluster data. The isochrones
used by \citet{an07} define a ratio, $\Delta M_V/\Delta$[Fe/H], over the color range of interest that is typically 1.4,
significantly larger than our derived value near 1. With the Hyades CMD position fixed via parallax, differential shifts with
metallicity then define the predicted position of clusters like M67. Even for a change in [Fe/H] of 0.13, an overestimate of
25$\%$ in the slope would lead to an underestimate of the distance by 0.03 mag.

Finally, at the extreme end of the age and metallicity scales among open clusters, the distance to NGC 6791 is derived
using field stars of comparable metallicity. Initial comparisons using the adjusted Hyades relation exhibited evidence
for significant evolutionary effects off the main sequence for the bluer stars in our sample, implying that the
average star in the comparison is older than the Hyades.  This trend virtually disappears when the data are compared to
the fiducial relation for the cluster. In fact, the bluest stars in the field are systematically fainter than the 
stars in the cluster at the same color, indicating that they are, as expected, younger than NGC 6791. Adopting
[Fe/H] = 0.40 and E$(B-V)$ = 0.15, we find $(m-M)_0$ = 13.13 $\pm$ 0.09. By comparison, the definitive value tied
to analysis of the eclipsing binary, V20, using the same parameters is $(m-M)_0$ = 13.00 $\pm$ 0.10. The significance
of the difference is marginal, especially given the underlying question of the photometric zero-points in $V$ and
$B-V$.

\acknowledgements
We are exceptionally grateful to Dr. Deokkeun An for supplying a careful and valuable critique of the manuscript on short notice after
the prior referee's delay. Extensive use was made of the SIMBAD database, operating at CDS, 
Strasbourg, France and the WEBDA database maintained at the University of Vienna, Austria \linebreak
(http://www.univie.ac.at/webda). We are also grateful to the Astronomy Department at Indiana University for 
hospitality during our Fall 2008 stay when the core material discussed in this paper was revised and rewritten.

\clearpage
\figcaption[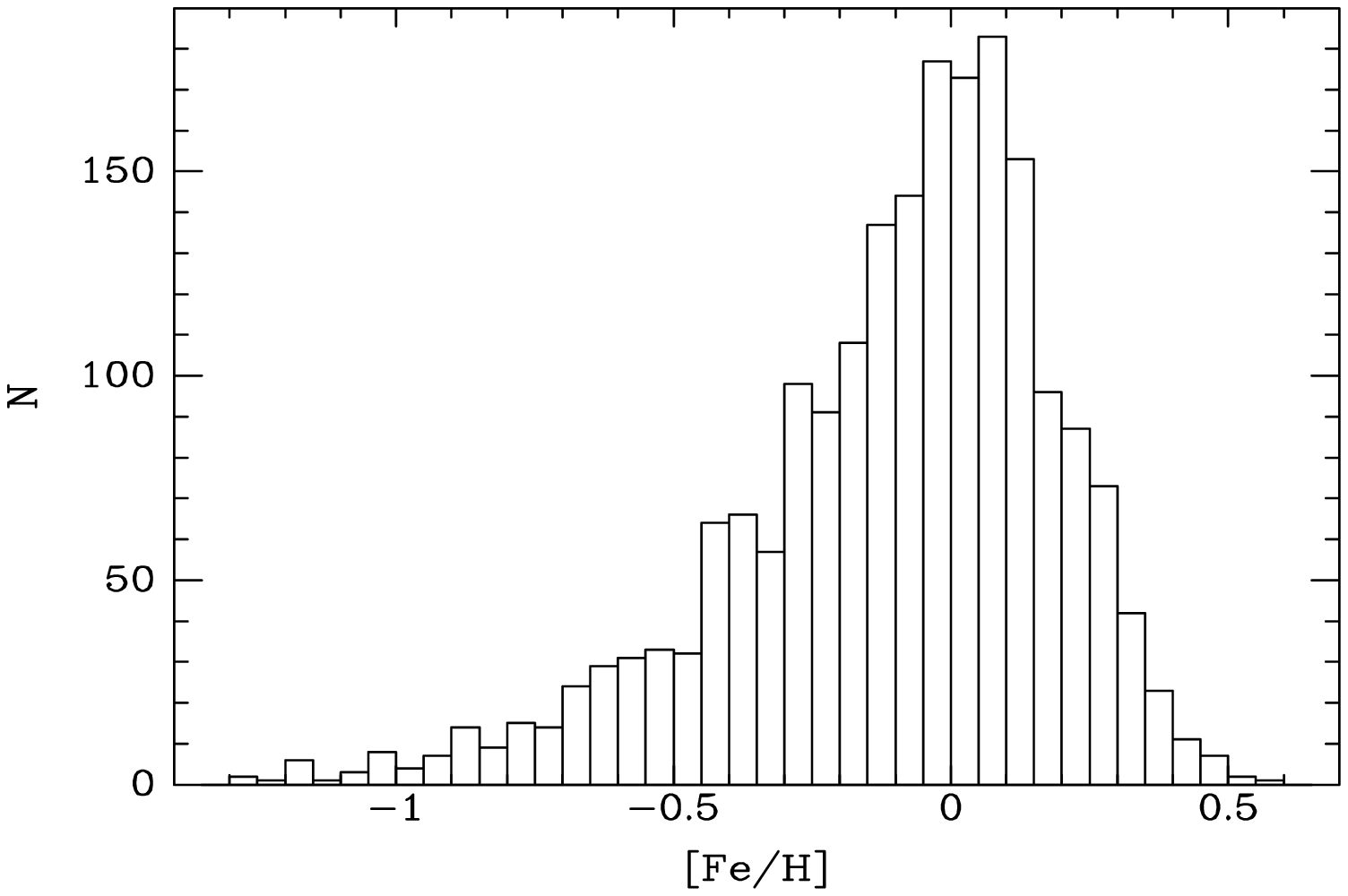]{Metallicity distribution for stars in the spectroscopic abundance catalog. \label{f1}}
\figcaption[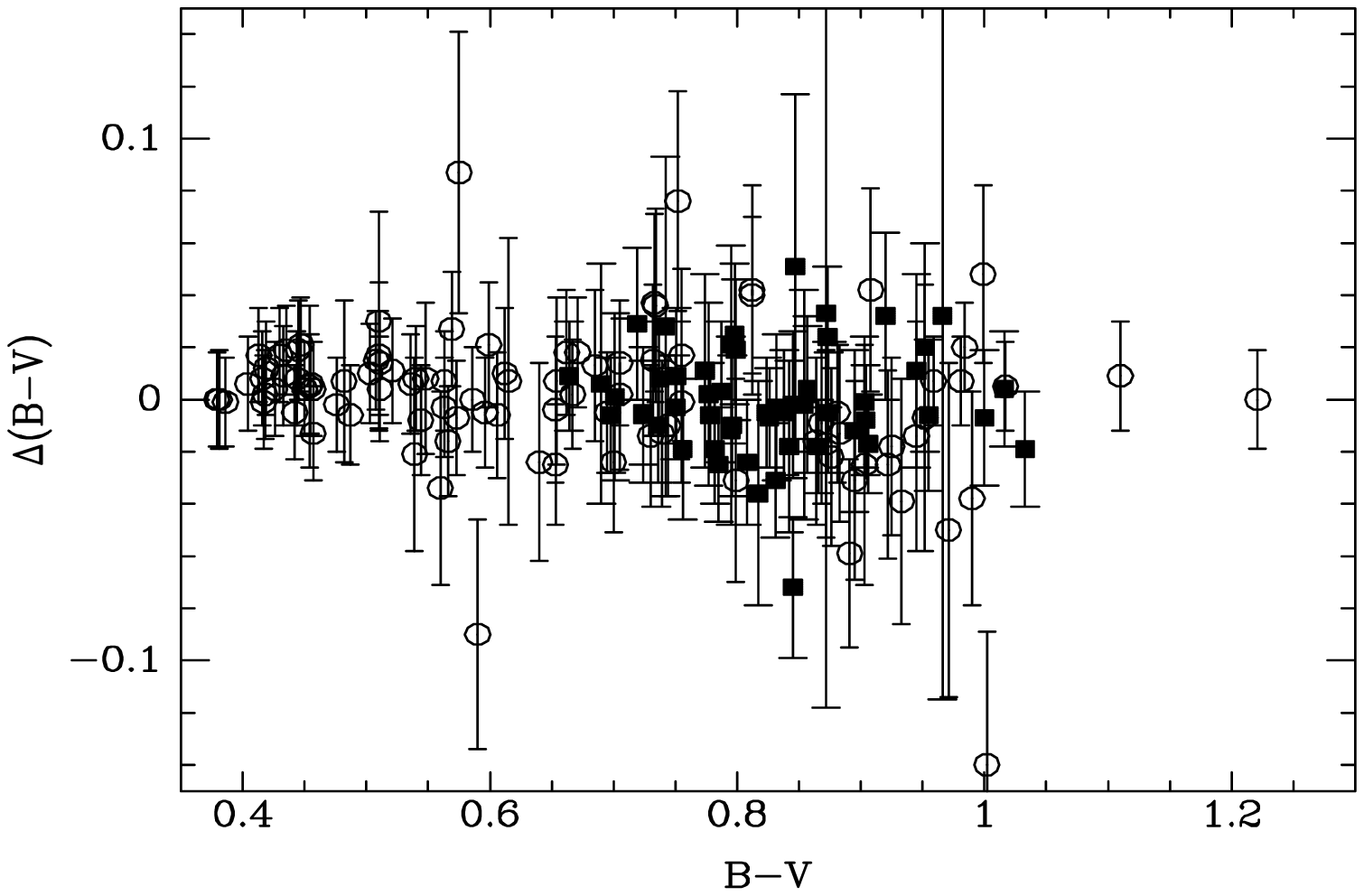]{Residuals in $B-V$, in the sense (LIT - converted Tycho), for \citet{pe03} (filled squares) and the Hyades photometry of 
\citet{joh55} (circles), adjusted by +0.011. \label{f2}} 
\figcaption[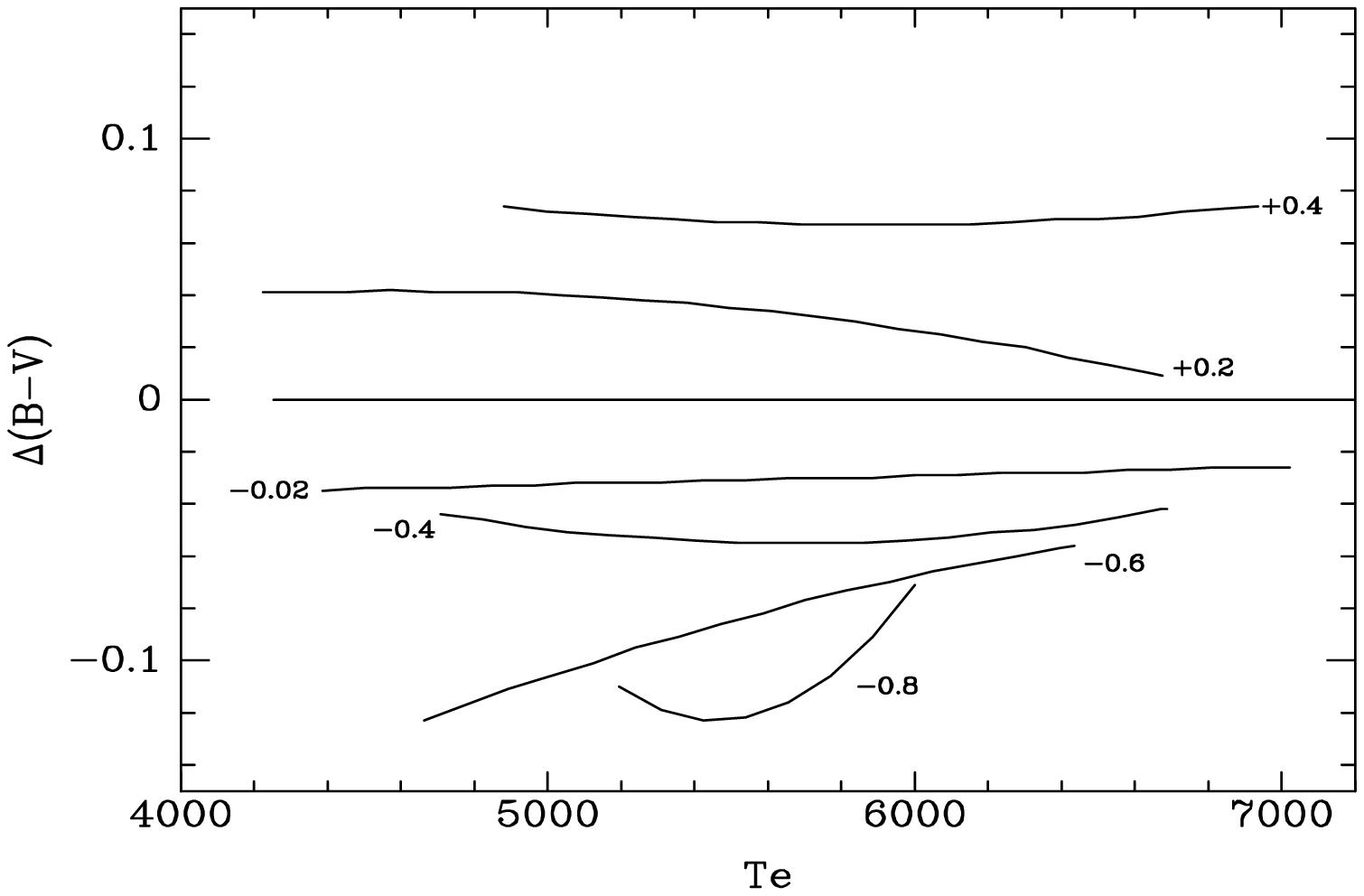]{Differences in $B-V$ relative to the solar relation as a function of effective temperature for stars of different 
metallicity. The [Fe/H] value for each relation is marked next to the relation. [Fe/H] = 0 is, by definition, the horizontal relation at 0. 
Temperature ranges for each curve are set by the sample limits in each metallicity range. \label{f3}}
\figcaption[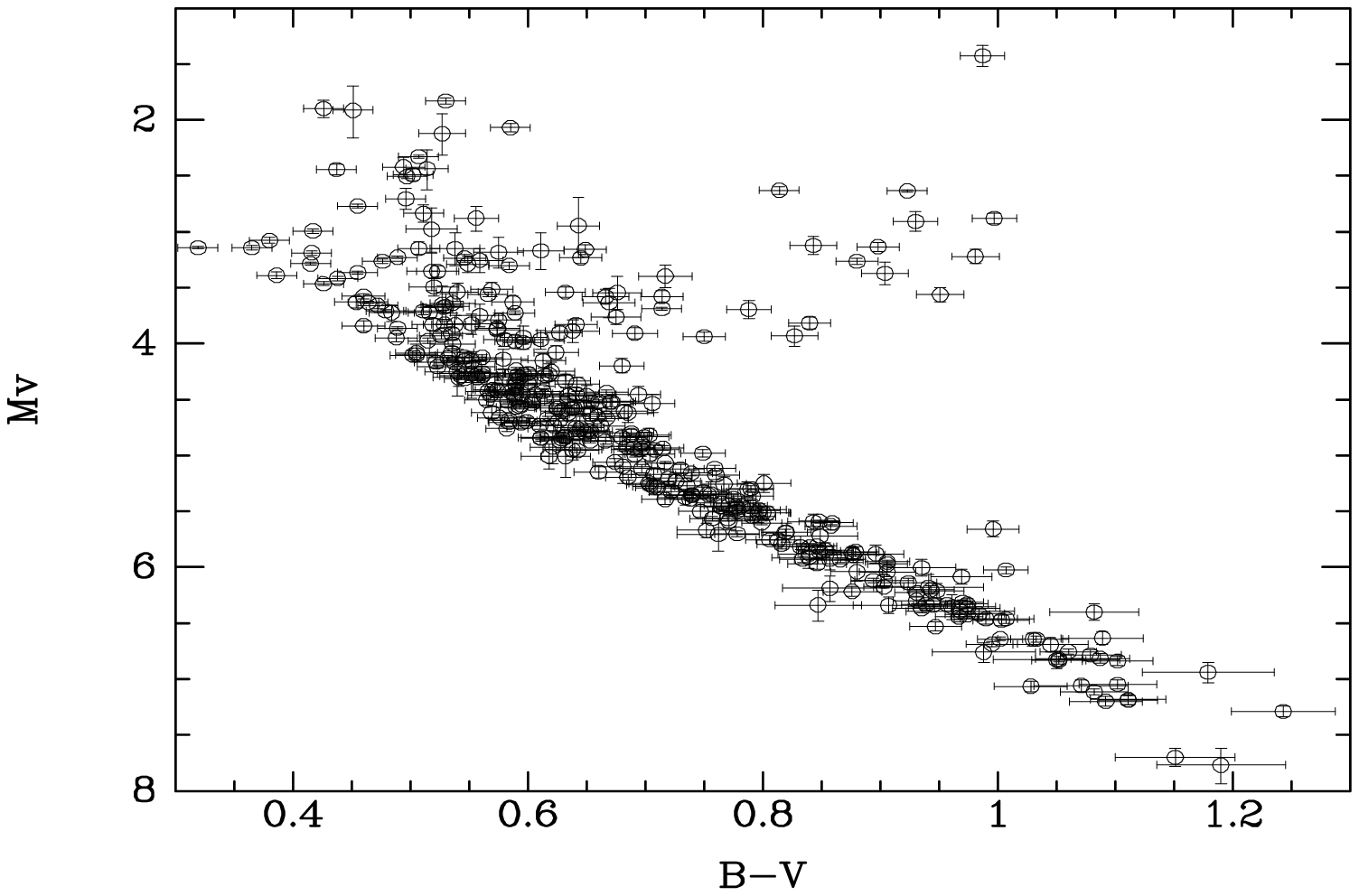]{CMD for stars between [Fe/H] = +0.05 and $-0.05$. Errorbars are based upon the errors in the parallaxes and the quoted errors
in the photometry. \label{f4}}
\figcaption[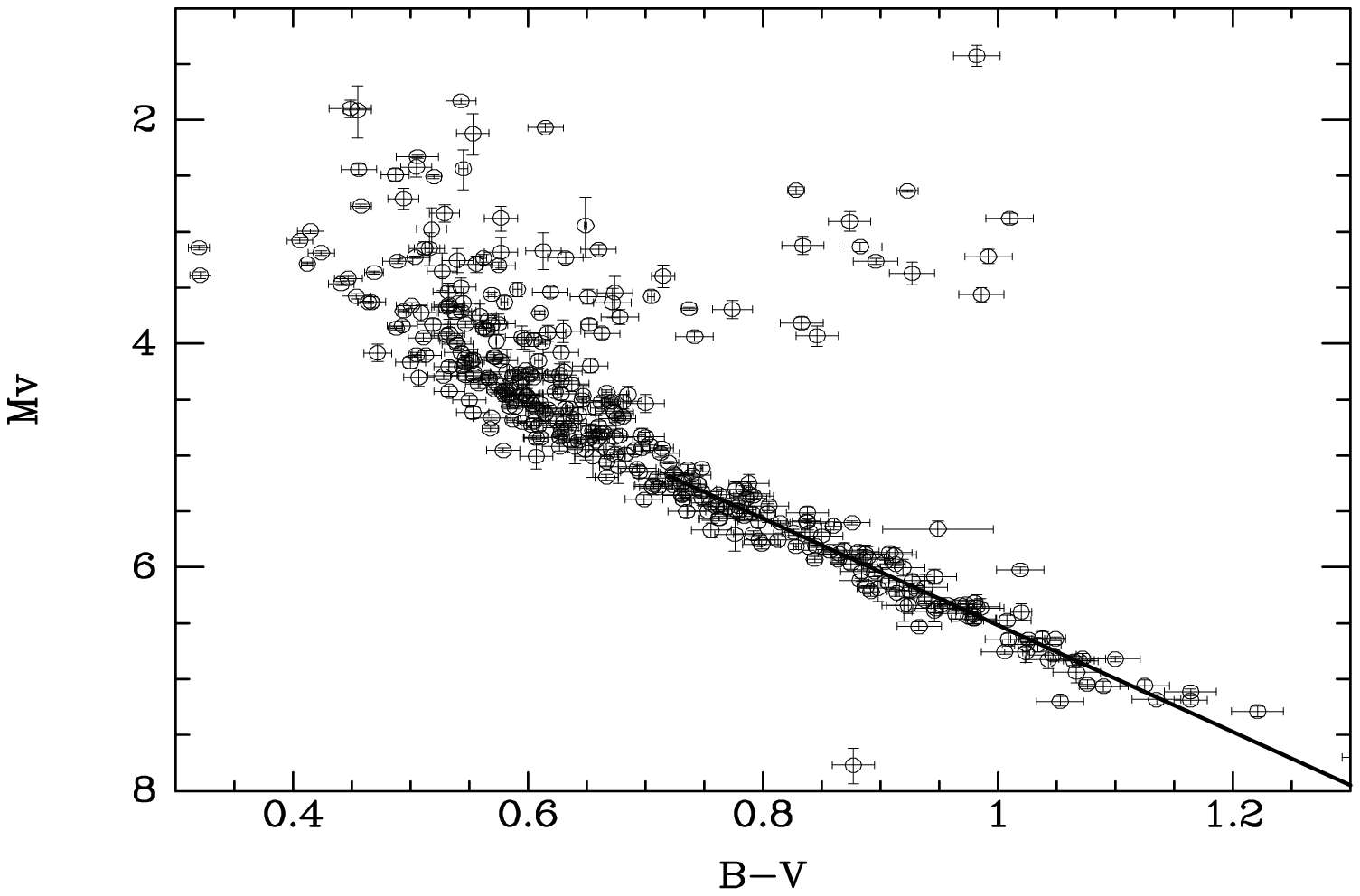]{Same as Fig. 4 with the $B-V$ colors derived using the transformation from $T_e$. \label{f5}}
\figcaption[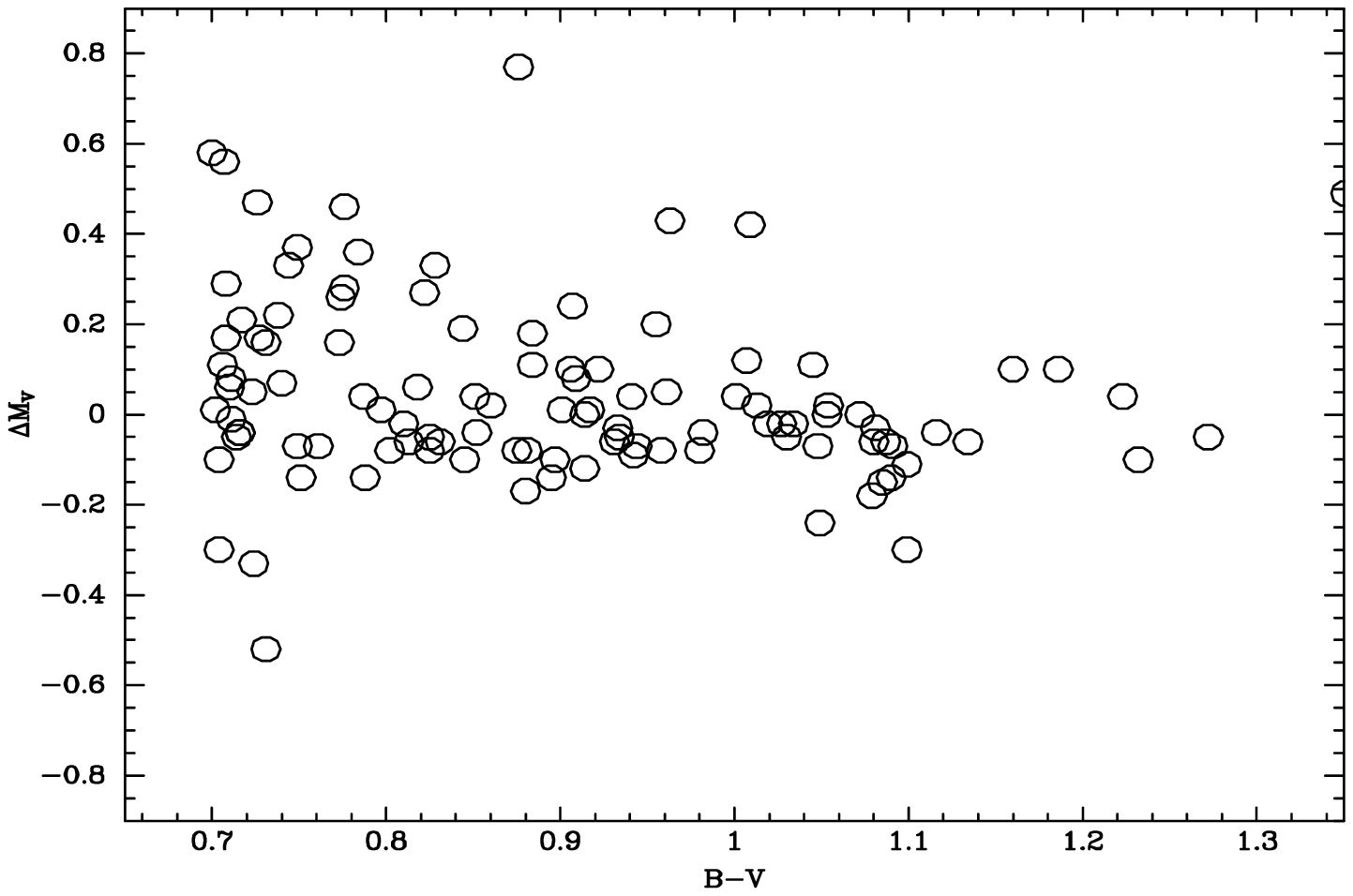]{The residuals, in the sense $(M_{VHy} - M_{VField})$, as a function of $B-V$ for parallax stars with [Fe/H]
between +0.08 and +0.18, adjusted in $M_V$ to [Fe/H] = +0.13 using $\Delta M_V/\Delta$[Fe/H]$ = 1.04$. \label{f6}}
\figcaption[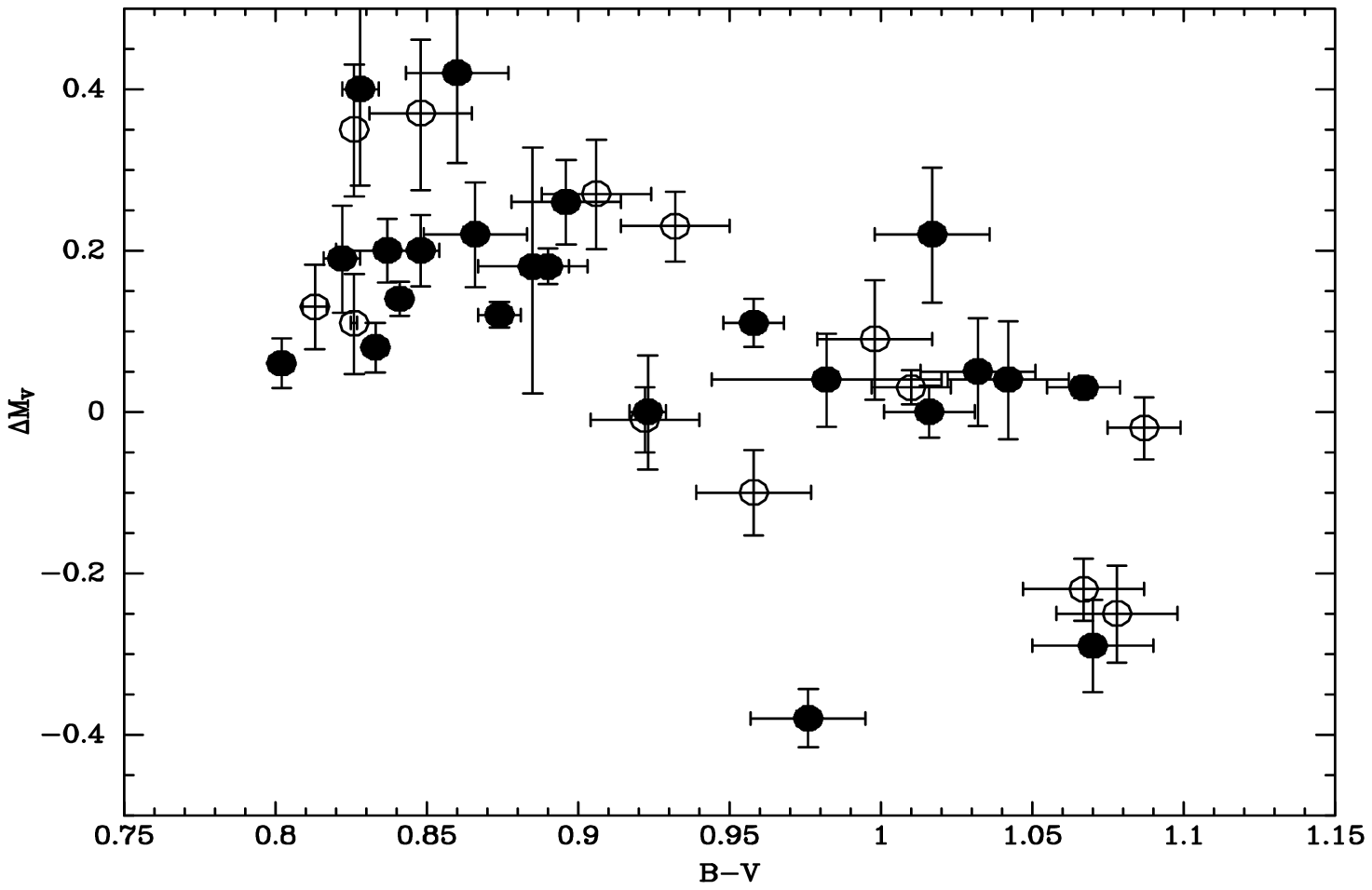]{The residuals between the field star data, adjusted to [Fe/H] = 0.39, and the shifted Hyades relation 
using $\Delta M_V/\Delta$[Fe/H]$ = 0.98$. Filled circles are stars with [Fe/H] above 0.29, while open circles are stars 
between [Fe/H] = 0.25 and 0.29.\label{f7}}
\figcaption[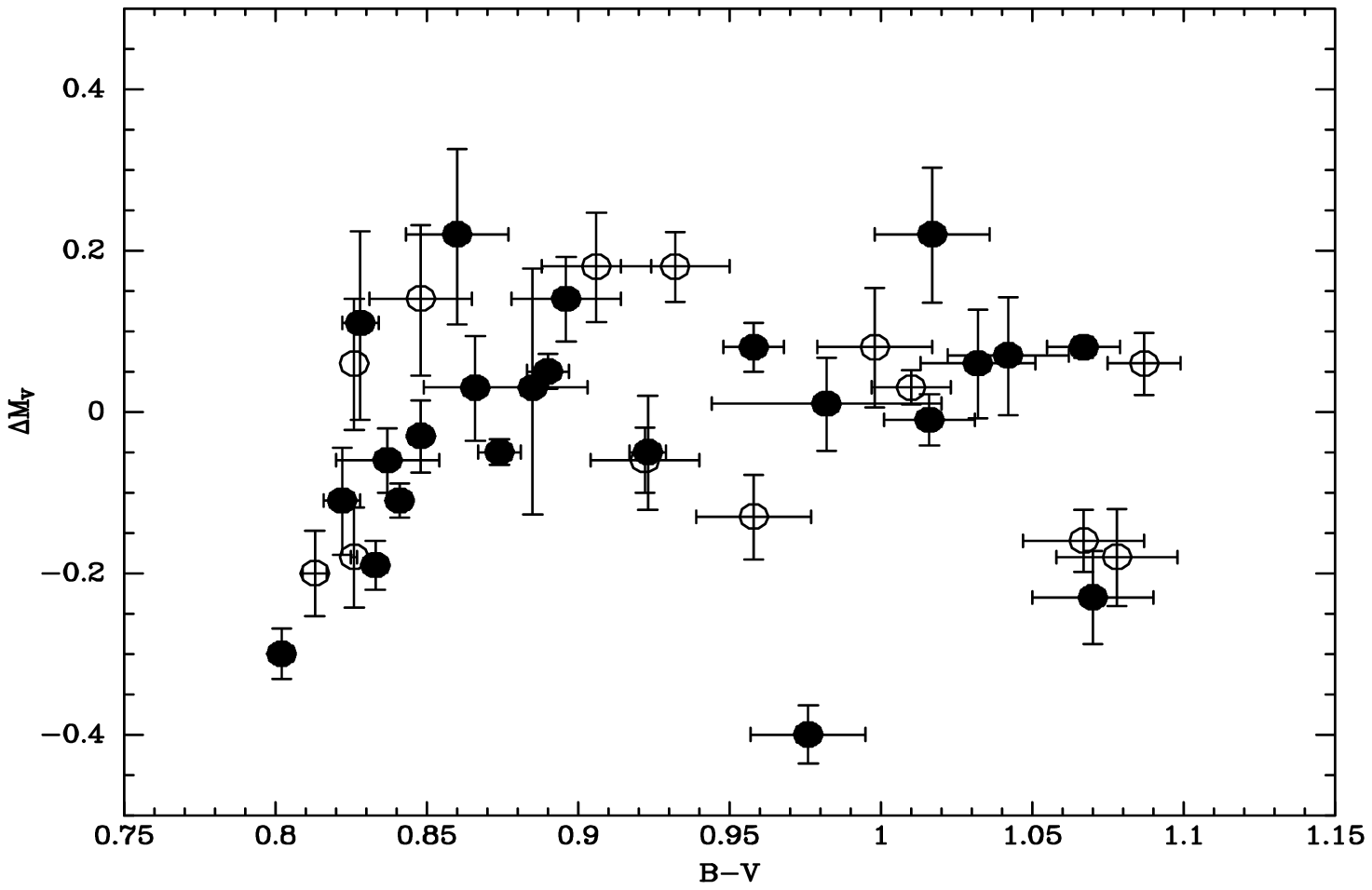]{Same as Fig. 7 using the fiducial relation of NGC 6791 for the comparison. E$(B-V)$ = 0.125 and $(m-M)$ = 13.45
have been adopted for the cluster parameters. \label{f8}}
\newpage
\plotone{f1.eps}
\plotone{f2.eps}
\plotone{f3.eps}
\plotone{f4.eps}
\plotone{f5.eps}
\plotone{f6.eps}
\plotone{f7.eps}
\plotone{f8.eps}
\end{document}